\documentclass{amsart}

\theoremstyle{definition}

\theoremstyle{remark}

\numberwithin{equation}{section}

\begin{document}

\title[Density of eigenvalues in unitary ensembles]
{Density of eigenvalues and its perturbation invariance
 in unitary ensembles of random matrices}

\author{Dang-Zheng Liu, Zheng-Dong Wang}
\address{SCHOOL OF MATHEMATICAL SCIENCES, PEKING UNIVERSITY,
BEIJING, 100871, P. R. CHINA}
\email{DZLIUMATH@GMAIL.COM}

\address{SCHOOL OF MATHEMATICAL SCIENCES, PEKING UNIVERSITY,
BEIJING, 100871, P. R. CHINA}

\email{ZDWANG@PKU.EDU.CN}
\author{Kui-Hua Yan}
\address{SCHOOL OF MATHEMATICS AND PHYSICS, ZHEJIANG NORMAL
UNIVERSITY, ZHEJIANG JINHUA, 321004, P. R. CHINA}
\email{YANKH@ZJNU.CN}


\begin{abstract}
We generally study the density of eigenvalues in unitary ensembles
of random matrices from the recurrence coefficients with regularly
varying conditions for the orthogonal polynomials. First we
calculate directly the moments of the density. Then,
 by studying some deformation of the moments, we get a family of differential equations of first order
 which the densities satisfy (see Theorem 1.2), and give the densities by solving them.
 Further, we prove that the density is invariant
after the polynomial perturbation of the weight function (see
Theorem 1.5).
\end{abstract}

\maketitle



\section{Introduction and main results}

Statistical behavior of eigenvalues of random matrices was first
studied by E.P. Wigner in 1950s to get an information about spectra
of heavy nuclei, and the famous semicircle-law was found then. In
1960s random matrices were intensively developed by E.P. Wigner,
F.J. Dyson, M.L. Mehta and others for a better understanding of
statistical behavior of energy levels in nuclear Physics, see [Po]
as a collection of early papers. Later more and more importance was
gained in other areas of Physics and Mathematics (see [Me] for a
comprehensive introduction, or J. Phys. A 36, 2003).

It is of special importance to observe that the eigenvalue
distribution of unitary invariant ensembles which can be described
in [FK] by weight functions from the families of classical
orthogonal polynomials. Such ensembles can be defined by the
probability distribution on a space $H_N$ of $N$-order Hermitian
matrices as

$$P(d A)=F(A)\prod_{j=1}^{N}d a_{jj}\prod_{j<k}d\Re a_{jk}d\Im a_{jk} \eqno (1.1)$$
where $F$ is an integrable nonnegative class function on $H_N$.

Using Proposition 2 in [HZ] we obtain

{\begin{eqnarray*} &&\hspace*{-1.0cm}\int_{H_{N}}F(A)d A=\\
&&\hspace*{-.5cm}Z_{N}\displaystyle\int^{+\infty}_{-\infty}\displaystyle\cdots\int^{+\infty}_{-\infty}
F{(\mbox{diag}{(x_{1},\ldots, x_{N})})} \prod_{1\leq j<k \leq N}\
(x_{j}-x_{k})^{2} \ d x_{1} \ldots d x_{N}  \hspace*{0.5cm} \ (1.2)
\end{eqnarray*}}where\ $$Z_{N}=\frac{\pi^{k(k-1)/2}}{k! (k-1)!\cdots 1!}.\eqno
(1.3)$$ Especially when $$ F(\mbox{diag}{(x_{1},\ldots, x_{N})})=
Z_{N}^{-1}\prod_{j=1}^{N}\omega
 (x_{j})\chi_{I}(x_{j}), \eqno (1.4)$$ one introduces the joint distribution
 function for the eigenvalues

 $$P_{N}(x_{1}, x_{2}, \ldots, x_{N})=\prod_{j=1}^{N}\omega
 (x_{j})\prod_{1\leq j<k \leq N}\ (x_{j}-x_{k})^{2} \eqno (1.5)$$
 where $I=(\alpha, \beta),\ -\infty\leq \alpha<\beta \leq
 +\infty$ and  $\omega
 (x)$ is a weight function on the interval $I$ all finite
 moments of which exist.

\vskip 3mm
 \noindent Remark 1.1 \ (1.5) was imposed on a definition of a matrix ensemble in [Me], and obtained in [TW] for
  $$F(A)=\mbox{exp}\big(-\mbox{Tr}(V(A))\big)\eqno (1.6)$$
   where $V(x)$ is a real-valued function such that $\omega
 (x)=\mbox{exp}(-V(x))$ defines a weight function. In addition,
 it is easy to draw the conclusion similar to (1.5) for the orthogonal and symplectic
 ensembles.

\vskip 3mm \noindent Remark 1.2 \ The appearance of the
characteristic function $\chi_{I}$ of the set $I$ means that we just
consider sub-ensemble of Hermitian matrices space whose eigenvalues
are all in $I$. Moreover, $I$ can be given by the union of some
disjoint intervals and then $\omega
 (x)$ is the associated weight function.

From (1.5) the $n-$point correlation function is defined in [Me] by

$$R_{n}{(x_{1},\ldots, x_{n})}=
\frac{N!}{(N-n)!}\displaystyle\int_{I}\displaystyle\cdots\int_{I}
P_{N}(x_{1}, x_{2}, \ldots, x_{N})\ d x_{n+1} \ldots d x_{N}. \eqno
(1.7)$$

If we introduce the orthogonal polynomials
$$\int_{I}p_{j}(x)p_{k}(x)\omega(x)dx=\delta_{jk},\ \ j,k=0,1,\ldots \eqno (1.8)$$
and the associated functions
$$\varphi_{k}(x) = p_{k}(x)\sqrt{\omega(x)} , \eqno
(1.9)$$ then by the property of Vandermonde determinant the joint
distribution function (1.5) reads:

 {\begin{eqnarray*}
 \hspace*{2.2cm} P_{N}(x_{1}, x_{2}, \ldots, x_{N})\hspace*{-2.2mm}&=&\hspace*{-2.2mm}\frac{1}{N!}
 \big(\mbox{det}[\varphi_{j-1}(x_{k})]|_{j,k=1}^{N}\big)^{2}\\
  \hspace*{-2.2mm}&=&\hspace*{-2.2mm}\frac{1}{N!} \mbox{det}[K_{N}(x_{j},x_{k})]|_{j,k=1}^{N}
  \hspace*{2.2cm}\ \ (1.10)
 \end{eqnarray*}}where
$$K_{N}(x,y)=\sum_{j=0}^{N-1}\varphi_{j}(x)\varphi_{j}(y).\eqno (1.11)$$
Because of the orthonormality of the $\varphi_{j}(x)$'s one can show
just as in [Me, Chap.6] that
$$R_{n}{(x_{1},\ldots, x_{n})}=
\mbox{det} [K_{N}(x_{j},x_{k})]|_{j,k=1}^{n}. \eqno (1.12)$$

It is of considerable interest to obtain the behavior of $n-$point
correlation of (1.12) after some appropriate scaling in the limit of
large $N$ . In particular, putting $n=1$, we get the density of
eigenvalues (also called level density )
$$R_{1}(x)=K_{N}(x,x)=\sum_{j=0}^{N-1}\varphi^{2}_{j}(x) \eqno
(1.13)$$ and denote the normalized density by
$$\frac{1}{N}R_{1}(x)=\frac{1}{N}\sum_{j=0}^{N-1}\varphi^{2}_{j}(x).\eqno (1.14)$$

\vskip 3mm Remark 1.3 \ For the eigenvalues $x_{1}, x_{2}, \ldots,
x_{N}$ of a random Hermitian matrix $A$, the integral of the
normalized counting function over an interval $\Delta=(c,d)$ is
given by
$$\nu_{N}(\Delta)=\frac{1}{N} \int_{\Delta}\sum_{j=1}^{N}\delta(x-x_{j})dx . \eqno
(1.15)$$ A calculation in [PS] shows that
$$\mathbb{E}(\nu_{N})(\Delta)=\int_{\Delta}\frac{1}{N}R_{1}(x)dx \eqno
(1.16) $$ where $\mathbb{E}(\nu_{N})(\Delta)$ denotes the
expectation of $\nu_{N}(\Delta)$ with respect to the probability
measure of (1.5).

Our motivation is to obtain the density in the scaling limit of
large $N$ from (1.14) and further study its polynomial-perturbation
invariance. Before our results are stated we review some known
results about the density.

As we know,  Wigner in [Wig1, 2] not only got his famous semicircle
law (here corresponding to the weight $\omega(x)=e^{-x^{2}}$ called
Gauss unitary ensemble, denoting GUE)
$$\sigma(x)=\frac{2}{\pi} \sqrt{1-x^{2}} \ \chi_{[-1, 1]}(x),\eqno (1.17)$$
but also invented the calculation method  which had some independent
interest as he thought. Afterwards, Bronk got the associated density
for $\omega(x)=x^{\rho}e^{-x},\ \rho>-1, 0<x<+\infty$ for Laguerre
ensembles in [Br] and Leff for $\omega(x)=(1-x)^{a}(1+x)^{b},\
a,b>-1,\ -1<x<1$ for Jacobian ensembles in [Lef]. Recently, an
irradiative idea in [HT] was introduced by Haagerup and Thorbjornsen
 to give a  short proof of Wigner's semicircle law, using the Laplace
transform of (1.14) and the property of Hermitian polynomials. In
[Lef], Ledoux pushed forward the investigation in [HT] and in a
slightly different way obtained Wigner's semicircle law, also the
densities for  Laguerre and Jacobian unitary ensembles but whose
expressions appeared complicated. In addition, based on the spirit
of statistical mechanics the method of equilibrium measure is used
to obtain the density, mainly for the weight function
$\omega(x)=\mbox{exp}(-V(x))$,
$$V(x)=\gamma_{2m}x^{2m}+\cdots+\gamma_{0},\ \gamma_{2m}>0. \eqno
(1.18)$$ See [Jo], [DMK] or [De] for the equilibrium measure method.
In a recent survey on orthogonal polynomial ensembles ([K\"{o}]),
K\"{o}nig obtained semicircle law for GUE respectively by moment
method and equilibrium measure method.

In this paper, we will generally deal with the weight function
$\omega(x)$  on the interval $I$  and obtain the density of
eigenvalues $\sigma(x)$. In fact, Professor W. Van Assche told us
that P. Nevai and W. Van Assche had shown that the density and zero
distribution of the orthogonal polynomials were the same ([Va1],
Theorem 5.3), see Remark 1.8 below. However, our method is to
calculate directly the moments without other knowledge than
three-term recurrence formula , and further the density was obtained
using a different method. To our knowledge, the family of densities
satisfying differential equations of first order is first obtained
in the present paper.

It is a well known result that three-term recurrence formula holds
for the orthogonal polynomials defined by (1.8)
$$x p_{n}(x)=a_{n+1}p_{n+1}(x)+b_{n}p_{n}(x)+a_{n}p_{n-1}(x),\ n=0,1,\cdots \eqno (1.19)$$
where $ a_{n}>0, p_{-1}(x)=0$. We will assume that there is a
positive and non-decreasing sequence $c_{n}$ such that
$$\lim_{n\rightarrow \infty}\frac{a_{n}}{c_{n}}=a>0,\ \lim_{n\rightarrow \infty}\frac{b_{n}}{c_{n}}=
b. \eqno (1.20)$$ An extra condition on the contraction sequence is
that $c_{n}$ is a regularly varying sequence with index $\lambda\geq
0$, i.e.
$$c_{n}=n^{\lambda}L(n)\eqno (1.21)$$
where $L:(0,+\infty)\longrightarrow (0,+\infty)$ is slowly varying ,
that is, $$\lim_{x\rightarrow \infty}\frac{L(xt)}{L(x)}=1,\ \ \
\forall t>0. \eqno (1.22)$$ The condition of (1.21) was first
introduced by W. Van Assche to study the asymptotics for orthogonal
polynomials, see [Va2] as a general survey for the condition of
(1.20) and [Fe, Chap.VIII] for regular functions (1.22).

\vskip 3mm Remark 1.4 \ Assuming $c(x)$ is a positive,
 non-decreasing differentiable  function on the interval
$(0,+\infty)$ with
$$\lim_{x\rightarrow +\infty}\frac{x c^{'}(x)}{c(x)}=\lambda \geq 0, \eqno (1.23)$$
then one easily knows  that $c(x)$ can be represented as
$$c(x)=x^{\lambda}L(x). \eqno (1.24)$$
A discrete version of this result is that (1.21) holds if
$$\lim_{n\rightarrow \infty}n \Big(\frac{c_{n+1}}{c_{n}}-1\Big)=\lambda \geq 0. \eqno (1.25)$$

\vskip 3mm Remark 1.5 \ The conditions of (1.20) and (1.21) are
connected close with the asymptotic problems of the orthogonal
polynomials. The class of Freud weights plays a most close role on
(1.21), for examples, $\omega(x)=\mbox{exp}(-Q(x))$ where $Q(x)$
grows like a power at infinity, in particular,
$$Q(x)=\gamma_{2m}x^{2m}+\cdots+\gamma_{0},\
\gamma_{2m}>0\eqno(1.26)$$
 with
$\lambda=1/(2m)$, see [DKMVZ]
;$$\omega(x)=|x|^{\beta}e^{-|x|^{\alpha}},\ \beta>-1,\ \alpha>0
\eqno(1.27)$$ with $\lambda=1/\alpha$, see [LMS] or [Va1]. In
addition, $\lambda=1$ for Laguerre weights
$$\omega(x)=x^{\alpha}e^{-x},\ \alpha>-1 \eqno (1.28)$$
and $\lambda=0$ for Jacobi weights
$$\omega(x)=(1-x)^{\alpha}(1+x)^{\beta},\ \alpha, \beta>-1.\eqno(1.29)$$
A classic result of E.A. Rakhmanov in [Si] asserts that if the
weight $\omega$ on $[-1,1]$ satisfying $\omega>0$ a.e. on $[-1,1]$,
then $\omega (x)$ belongs to the Nevai-Blumenthal class, that is
 $$\lim_{n\rightarrow \infty}a_{n}=1/2,\ \lim_{n\rightarrow \infty}b_{n}=0. \eqno(1.30)$$
Obviously, $\lambda=0$ for the Nevai-Blumenthal class (see [Va2]).
We strongly refer the reader to [Lub] for a recent survey for a wide
variety of weights on finite or infinite intervals.

Now we can state our main results. In Section 2 we will introduce
ascending, equilibrating and descending operators which describe the
transforming of polynomials, and explicitly calculate the moments of
the density. Then in Section 3 we consider a simple deformation of
the moments for any given density and obtain a corresponding density
determined by a differential equation with respect to the new
moments. In section 4 using the results in Section 3 we give the
proofs of Theorems 1.2 and 1.5 below.

 First let us rescale the density of (1.14) by
$$\sigma_{N}(x)=\frac{c_{N}}{N}R_{1}(c_{N}x).\eqno (1.31)$$
Note that $\sigma_{N}(x)$ is our main object  and we will study its
limit behavior.

\vskip 0.5cm \noindent {\bf Theorem 1.1} \hskip 0.2true cm Denote
the $k\mbox{th}$ moment of the scaling density $\sigma_{N}(x)$ by
$M_{k}^{(N)}$. Under the contraction conditions of (1.20) and
(1.21), we have

$$\lim_{N\rightarrow \infty} M_{k}^{(N)}=M_{k},\ \ \ \ \ \ k= 0,1,\ldots \eqno (1.32)$$
where
$$M_{k}=\frac{1}{1+\lambda k}\Big (\sum_{j=0}^{[k/2]}C_{k}^{j} C_{k-j}^{j} a^{2j} b^{k-2j}\Big).\eqno (1.33)$$

Remark 1.6 \ One sets $a= 1/2$ in the following since $c_{n}$ can be
chosen freely from some constant. Observe
$$\sum_{j=0}^{[k/2]}C_{k}^{j} C_{k-j}^{j} a^{2j} b^{k-2j}=L_{0}\Big (a z+\frac{a}{z}+b\Big)^{k} \eqno (1.34)$$
where the operator $L_{0}(f)$ represents the constant term of
Laurent series, using Cauchy contour integral, that is
$$L_{0}(f)=\frac{1}{2\pi i} \oint \frac{f(z)}{z} d z.\eqno (1.35)$$

\vskip 0.5cm \noindent {\bf Theorem 1.2} \hskip 0.2true cm A
probability density $\sigma(x)$ with its $k\mbox{th}$ moment $M_{k}$
exists, and is uniquely determined by the following differential
equation of first order
$$\sigma(x)-\lambda \big[x\sigma(x)\big]^{(1)}=\frac{1}{\pi} \frac{1}{\sqrt{1-(x-b)^{2}}}\chi_{I_{b}}, \eqno (1.36)$$
with the following conditions
$$\sigma(x)\geq 0,  \int_{-\infty}^{+\infty}\sigma(x) d x=1  \eqno (1.37)$$
where $I_{b}=(-1+b, 1+b)$ and $\chi_{I_{b}}$ is a characteristic
function of $I_{b}$. Exactly, the support of $\sigma(x)$ can be
restricted to a finite interval, that is, for $\lambda =0$
$$\sigma (x)=\frac{1}{\pi} \frac{1}{\sqrt{1-(x-b)^{2}}}\eqno (1.38)$$
while for $\lambda >0$
$$ \mbox{supp}(\sigma) = [B_{1}, B_{2}] \eqno (1.39)$$
where $$B_{1}= \mbox{min}\{b-1, 0\}, B_{2}= \mbox{max}\{b+1,
0\}.\eqno (1.40)$$

\vskip .3cm One directly solves the equation of (1.36) and easily
obtains

 \vskip 0.5cm \noindent
{\bf Corollary 1.3} \hskip 0.2true cm  For $b=0$ and $m \in
\mathbb{N}$, we have

(1) for $\lambda =1/2m$
 $$\sigma(x)=\frac{4}{\pi}\sqrt{1-x^{2}}\sum_{j=1}^{m}\big((2x)^{2m-2j}C_{2j-2}^{j-1}\big)/C_{2m}^{m} \eqno (1.41)$$
and

(2) for $\lambda =1/(2m-1)$
$$\sigma(x)=\frac{m C_{2m}^{m}}{2 \pi} \Big((\frac{x}{2})^{2m-2} \ln \frac{1+\sqrt{1-x^{2}}}{|x|}+
\frac{\sqrt{1-x^{2}}}{2}\sum_{j=1}^{m-1}\frac{1}{j
C_{2j}^{j}}(\frac{x}{2})^{2m-2-2j} \Big). \eqno (1.42)$$

For $b=-1$ and $q \in \mathbb{N} \cup \{0\}$, we have

(3) for $\lambda =1/(q+1)$
 $$\sigma(x)=\frac{q+1}{\pi} \big(\frac{x}{2}\big)^{q} \sum_{j=0}^{q}\frac{C_{q}^{j}}{1+2j}
 \Big(\sqrt{\frac{2-x}{x}}\Big)^{1+2j}
  \eqno (1.43)$$
and

(4) for $\lambda =1/(q+\frac{1}{2})$
$$\sigma(x)=\frac{2q+1}{4 \pi} \Big((\frac{x}{8})^{q-\frac{1}{2}} \ln \big (\sqrt{\frac{2-x}{x}}+\sqrt{\frac{2}{x}}\big)
+\sqrt{\frac{2-x}{x}}\sum_{j=1}^{q}\frac{1}{j
C_{2j}^{j}}(\frac{x}{8})^{q-j} \Big) C_{2q}^{q}. \eqno (1.44)$$


Remark 1.7 \ For the weight $\omega(x)=\mbox{exp}(-V(x))$ from
(1.18) which is corresponding with $\lambda =1/2m$, the density was
given in [Jo] in the form of $r(x)\sqrt{(x-x_{1})(x_{2} -x)}$. Here
$x_{1}<x_{2}$ and $r(x)$ is a polynomial of degree $2m-2$ depending
on $V(x)$; when $\lambda=1$ and $b=0$, it was obtained in [CI] using
Mathematica for Meixner-Pollaczek polynomials. It is just the case
 where $\lambda=1$ and $b=-1$ for Laguerre weights of (1.28), see [Sze].
 The authors  believe that
the densities in Corollary 1.3, especially for small integrals $m$
and $q$, may appear in some other random matrix models.

\vskip .3cm
 From Theorems 1.1 and 1.2, it is obvious that

 \vskip 0.5cm \noindent {\bf Theorem 1.4} \hskip 0.2true cm Denote
 $\sigma_{N}(x)$ and $\sigma(x)$ as above , then
$$\sigma_{N}(x) \stackrel{\mathrm{W}}{\longrightarrow}\sigma(x) \eqno(1.45)$$
where $\mathrm{W}$ means in the weak sense.

\vskip .3cm
 Remark 1.8 \ Let $x_{1,n},
x_{2,n},\cdots, x_{n,n}$ be zeros of $p_{n}$. P. G. Nevai and J. S.
Dehesa in [ND] also got the same moments of zero distribution under
the contraction condition of (1.20) (this fact was pointed out by W.
Van Assche in [Va2]), namely,
$$\lim_{n\rightarrow \infty}\frac{1}{n} \sum_{j=1}^{n}(\frac{x_{j,n}}{c_{n}})^{k}= M_{k}. \eqno (1.46)$$
A beautiful probabilistic interpretation in [Va1] shows $M_{k}$ is
the $k$th moment of Nevai-Ullman measure. In addition, Nevai and Van
Assche proved that weak convergence of zero distribution implied
some weak convergence
 to the same probability measure for Christoffel functions, which are defined by $(\displaystyle\sum_{j=0}^{N-1} p^{2}_{j}(x))^{-1}$
 (see Theorem 5.3 in [Va1]). It is obvious that the density of eigenvalues and zero distribution are the same
 according to (1.46) and Theorem 1.1, thus we give a new proof.

\vskip .3cm

Let p(x) be a fixed $l$th order polynomial, and we consider a new
weight function
$$\hat{\omega}(x)=p^{2}(x)\omega(x)\eqno (1.47)$$
with
$$\int_{I}\hat{p}_{j}(x)\hat{p}_{k}(x)\hat{\omega}(x)dx=\delta_{jk},\ \
j,k=0,1,\ldots .\eqno (1.48)$$ Associated 1-point correlation
function is given by
$$\hat{R}_{1}(x)=\sum_{j=0}^{N-1}\hat{p}^{2}(x)p^{2}(x)\omega(x).\eqno (1.49)$$
Under the same scaling we write
$$\hat{\sigma}_{N}(x)=\frac{c_{N}}{N}\hat{R}_{1}(c_{N}x). \eqno (1.50)$$

Calculate the moments of (1.50) for the new weight function, we find

\vskip 0.5cm \noindent {\bf Theorem 1.5} \hskip 0.2true cm Denote
the $k\mbox{th}$ moments of $\hat{\sigma}_{N}(x)$ and
$\sigma_{N}(x)$ by $M_{k}^{(N)}$ and $\hat{M}_{k}^{(N)}$,
respectively. Under the contraction conditions of (1.20) and (1.21),
we have

$$\lim_{n\rightarrow \infty} \hat{M}_{k}^{(N)}=\lim_{n\rightarrow \infty} M_{k}^{(N)}
=M_{k},\ \ \ \ \ k= 0,1,\ldots. \eqno (1.51)$$ Namely,
$$\hat{\sigma}_{N}(x) \stackrel{\mathrm{W}}{\longrightarrow}\sigma(x) \eqno(1.52)$$
where $\sigma(x)$ is the density determined by (1.36) and (1.37).

\section{Calculation of the moments}

By the recursion formula of (1.19), we regard the multiplication by
x as an operator $A_{x}$, and it can be represented as
$$A_{x}=A_{+}+A_{0}+A_{-} \eqno (2.1)$$
where $A_{+}, A_{0} \mbox{\ and} \  A_{-}$  are called ascending,
equilibrating and descending operators respectively, defined by
$$A_{+}p_{n}(x)= a_{n+1}p_{n+1}(x),\ \ \
A_{0}p_{n}(x)= b_{n}p_{n}(x),\ \ \  A_{-}p_{n}(x)= a_{n}p_{n-1}(x).
\eqno (2.2)
$$
Thus we calculate the $k$th moment of $\sigma_{N}(x)$ using (1.31),
(2.1) and (2.2) as follows:

 {\begin{eqnarray*}
 \hspace*{1.8cm} M_{k}^{(N)}\hspace*{-1.5mm}&=&\hspace*{-1.5mm}\int x^{k}\sigma_{N}(x)d x\\
  \hspace*{-1.5mm}&=&\hspace*{-1.5mm}\frac{1}{N\ (c_{N})^{k}}\sum_{j=0}^{N-1}\int_{I} x^{k} p^{2}_{j}\omega(x)d
  x\\
\hspace*{-1.5mm}&=&\hspace*{-1.5mm}\frac{1}{N\
(c_{N})^{k}}\sum_{j=0}^{N-1}\Big\langle x^{k}
p_{j},p_{j}\Big\rangle_{L^{2}(\omega)}\\
\hspace*{-1.5mm}&=&\hspace*{-1.5mm} \frac{1}{N\
(c_{N})^{k}}\sum_{j=0}^{N-1}\Big\langle (A_{+}+A_{0}+A_{-})^{k}
p_{j},p_{j}\Big\rangle_{L^{2}(\omega)}.\hspace*{1.3cm}\ \ (2.3)
 \end{eqnarray*}}

Let $\Lambda_{k}^{q}$ be a set composed of those terms in the
expansion of $(A_{+}+A_{0}+A_{-})^{k}$, in which the operators
$A_{+}$ and $A_{-}$ exactly appear $q$ times respectively. Note that
$\big\langle T p_{j},p_{j}\big\rangle_{L^{2}(\omega)}= 0 \
\mbox{for}\  T\not\in \bigcup\limits_{q} \Lambda_{k}^{q}$, then one
obtains the following lemma:

\vskip 0.5cm \noindent {\bf Lemma 2.1} \hskip 0.2true cm
$$M_{k}^{(N)}=\frac{1}{N\
(c_{N})^{k}}\sum_{j=0}^{N-1}\sum_{q=0}^{[k/2]}\sum_{T\in
\Lambda_{k}^{q} }\Big\langle T
p_{j},p_{j}\Big\rangle_{L^{2}(\omega)}.\eqno (2.4)$$

For convenience of the following calculation, we introduce one
property of the regular varying functions.

\vskip 0.5cm \noindent {\bf Lemma 2.2} \hskip 0.2true cm Let $c(x)$
be a positive and non-decreasing function on $(0,+\infty)$, and

$$c(x)=x^{\lambda}L(x)\eqno (2.5)$$
where $L(x)$ satisfies
 $$\lim_{x\rightarrow +\infty}\frac{L(xt)}{L(x)}=1,\ \ \
\forall t>0.\eqno (2.6)$$ Then

$$\lim_{n \rightarrow \infty}\frac{\frac{1}{n}\int_{1}^{n}(c(x))^{k} d x}{(c(n))^{k}}=\frac{1}{1+\lambda k}. \eqno (2.7)$$

\vskip 0.5cm \noindent {\bf Proof} \ By Lebesgue's dominated theorem
and exchanging limits and integrals , it is easy to prove using
(2.5) and (2.6). $\square$

\vskip .3cm
 Now we give a proof of Theorem 1.1 by two cases of $b$.

\vskip 0.3cm \noindent {\bf Case 1} \ $b\neq 0$. \vskip 0.3cm

Assume that $b> 0$ for convenience. Write
$$\frac{a_{n}}{c_{n}}=a (1+\xi_{n}),\ \ \frac{b_{n}}{c_{n}}=b (1+\eta_{n}). \eqno (2.8)$$
If one writes for  $j>k$
$$u_{j}=\max\limits_{j-k\leq m \leq j+k}\{|\xi_{m}|,\ |\eta_{m}|\}, \eqno (2.9)$$
then
$$\lim_{j \rightarrow \infty}u_{j}= 0. \eqno (2.10)$$
Thus we can suppose $u_{j}< 1,\forall j=0,1,\ldots$ and by the
definition of $u_{j}$ one obtains
$$\max\limits_{j-k\leq m \leq j+k}\{a_{m}\} \leq a\  c_{j+k}(1+u_{j}),\ \  \max\limits_{j-k\leq m \leq j+k}\{b_{m}\}
\leq b \ c_{j+k}(1+u_{j}).\eqno (2.11)$$ Furthermore, for $T\in
\Lambda_{k}^{q}$ we have
 {\begin{eqnarray*}
 \hspace*{.5cm} \Big\langle T p_{j},p_{j}\Big\rangle_{L^{2}(\omega)}\hspace*{-1.5mm}&\leq&\hspace*{-1.5mm}
 \Big(\max\limits_{j-k\leq m \leq j+k}\{a_{m}\}\Big)^{2q}\Big(\max\limits_{j-k\leq m \leq j+k}\{b_{m}\}\Big)^{k-2q}\\
  \hspace*{-1.5mm}&\leq&\hspace*{-1.5mm} a^{2q} b^{k-2q}(c_{j+k})^{k}(1+u_{j})^{k}.
  \hspace*{4.2cm}\ \ (2.12)
 \end{eqnarray*}}Summing by $q$ and $j$, we get
$$M_{k}^{(N)}\leq \Big (\sum_{q=0}^{[k/2]}C_{k}^{q} C_{k-q}^{q} a^{2q} b^{k-2q}\Big)
\frac{\frac{1}{N}\sum\limits_{j=0}^{N-1}(c_{j+k})^{k}(1+u_{j})^{k}}{(c_{N})^{k}}.\eqno
(2.13)$$ By Cauchy-Maclaurin summation formula and Lemma 2.2 one
obtains

{\begin{eqnarray*}
 \hspace*{1.3cm} \lim_{N\rightarrow \infty}\frac{\frac{1}{N}\sum\limits_{j=0}^{N-1}(c_{j+k})^{k}(1+u_{j})^{k}}{(c_{N})^{k}}
 \hspace*{-1.5mm}&=&\hspace*{-1.5mm}
 \lim_{N\rightarrow \infty}\frac{\frac{1}{N}\sum\limits_{j=0}^{N-1}(c_{j+k})^{k}}{(c_{N})^{k}}\\
  \hspace*{-1.5mm}&=&\hspace*{-1.5mm}
  \lim_{N\rightarrow \infty}\frac{\frac{1}{N}\sum\limits_{j=1}^{N}(c_{j})^{k}}{(c_{N})^{k}}\\
\hspace*{-1.5mm}&=&\hspace*{-1.5mm}
\lim_{N \rightarrow \infty}\frac{\frac{1}{N}\int_{1}^{N}(c(x))^{k} d x}{(c(N))^{k}}\\
\hspace*{-1.5mm}&=&\hspace*{-1.5mm} \frac{1}{1+\lambda k}.
\hspace*{3.9cm} \ (2.14)
 \end{eqnarray*}}

Analogously, for $j>k$, one obtains
$$M_{k}^{(N)}\geq \Big (\sum_{q=0}^{[k/2]}C_{k}^{q} C_{k-q}^{q} a^{2q} b^{k-2q}\Big)
\frac{\frac{1}{N}\sum\limits_{j=k}^{N-1}(c_{j-k})^{k}(1-u_{j})^{k}}{(c_{N})^{k}}\eqno
(2.15)$$ and
$$\lim_{N\rightarrow \infty}\frac{\frac{1}{N}\sum\limits_{j=k}^{N-1}(c_{j-k})^{k}(1-u_{j})^{k}}{(c_{N})^{k}}=
\frac{1}{1+\lambda k}. \eqno (2.16)$$

Combining (2.13) --- (2.16) we complete the proof of this case.
$\square$

 \vskip 0.3cm \noindent {\bf Case 2} \ $b=0$.
\vskip 0.3cm Similarly, write
$$\frac{a_{n}}{c_{n}}=a (1+\xi_{n}),\ \ \frac{b_{n}}{c_{n}}=\eta_{n} \eqno (2.17)$$
and
$$v_{j}=\max\limits_{j-k\leq m \leq j+k}\{|\xi_{m}|,\ |\eta_{m}|\}, \eqno (2.18)$$
then
$$\lim_{j \rightarrow \infty}v_{j}= 0. \eqno (2.19)$$
Thus we can suppose $v_{j}< 1,\forall j=0,1,\ldots$. For $0 \leq
j-k\leq m \leq j+k $, one obtains
$$a \ c_{j-k} (1-v_{j})\leq a_{m} \leq a\  c_{j+k}(1+v_{j}),\ \ |b_{m}|\leq c_{j+k} v_{j}. \eqno (2.20)$$

\vskip .3cm
 Now we give an estimation of $M_{k}^{(N)}$ when $k$ is odd. For $T\in
 \Lambda_{k}^{q}$, by (2.20)
we have
 {\begin{eqnarray*}
 \hspace*{2.1cm} \Big|\Big\langle T p_{j},p_{j}\Big\rangle_{L^{2}(\omega)}\Big|\hspace*{-1.5mm}&\leq&\hspace*{-1.5mm}
 a^{2q} v_{j}^{k-2q}(c_{j+k})^{k}(1+v_{j})^{k-2q}\\
  \hspace*{-1.5mm}&\leq&\hspace*{-1.5mm} a^{2q}
  v_{j}^{k-2q}(c_{j+k})^{k}(1+v_{j})^{k}.
  \hspace*{2.2cm}\ \ \ (2.21)
 \end{eqnarray*}}If denoting
$$\overline{v}_{j}=\sum_{q=0}^{[k/2]}C_{k}^{q} C_{k-q}^{q} a^{2q} v_{j}^{k-2q}, \eqno (2.22)$$
then
$$\lim_{j \rightarrow \infty}\overline{v}_{j}= 0. \eqno (2.23)$$
Thus by the Stoltz formula one obtains
$$\Big|M_{k}^{(N)}\Big|\leq \frac{\frac{1}{N}\sum\limits_{j=0}^{N-1}(c_{j+k})^{k}(1+v_{j})^{k}\overline{v}_{j}}{(c_{N})^{k}}
\longrightarrow 0. \eqno (2.24)$$

\vskip .3cm  When $k$ is even. For $T\in
 \Lambda_{k}^{q}, q \neq k/2$,
 $$\Big|\Big\langle T p_{j},p_{j}\Big\rangle_{L^{2}(\omega)}\Big| \leq
 a^{2q} v_{j}^{k-2q}(c_{j+k})^{k}(1+v_{j})^{k} \eqno (2.25)$$
 and for $T\in
 \Lambda_{k}^{k/2}, j>k$,
$$a^{k}(c_{j-k})^{k}(1-v_{j})^{k}
\leq\Big\langle T p_{j},p_{j}\Big\rangle_{L^{2}(\omega)}\leq
a^{k}(c_{j+k})^{k}(1+v_{j})^{k}.\eqno(2.26)$$ If one writes
$$\overline{M}_{k}^{(N)}=\frac{1}{N\
(c_{N})^{k}}\sum_{j=0}^{N-1}\sum_{q<k/2}\sum_{T\in \Lambda_{k}^{q}
}\Big\langle T p_{j},p_{j}\Big\rangle_{L^{2}(\omega)}\eqno (2.27)$$
and
$$\widetilde{M}_{k}^{(N)}=\frac{1}{N\
(c_{N})^{k}}\sum_{j=0}^{N-1}\sum_{T\in \Lambda_{k}^{k/2}
}\Big\langle T p_{j},p_{j}\Big\rangle_{L^{2}(\omega)},\eqno (2.28)$$
then
$$M_{k}^{(N)}=\overline{M}_{k}^{(N)}+\widetilde{M}_{k}^{(N)}. \eqno (2.29)$$

Similar to the case where $k$ is odd, by (2.25) one obtains
$$\lim_{N\rightarrow \infty}\overline{M}_{k}^{(N)}=0 \eqno (2.30)$$
and similar to the case where $b \neq 0$, by (2.26) one obtains
$$\lim_{N\rightarrow \infty}\widetilde{M}_{k}^{(N)}=
\frac{1}{1+\lambda k}C_{k}^{k/2}a^{k}. \eqno (2.31)$$

Combining  (2.24), (2.29), (2.30) and (2.31) we complete the proof
of this case where $b=0$. $\square$

Therefore, Theorem 1.1 has been proved.

\section{Deformation of the moments}

Following Wigner's original papers [Wig1] and [Wig2], to obtain the
distribution of eigenvalues in random matrix models, a standard
procedure is first to calculate the moments of all orders after some
appropriate scaling of eigenvalues and then to determine an explicit
distribution function of the moments. In this section we discuss
some deformations of the moments of a given density function and
determine the corresponding density function with respect to the new
moments, and then give an application in the following section. A
natural question: which kind of deformations of the moments has a
corresponding distribution function? We will make an interesting
try.

Let $f(x)$ be a continuous density function on the interval
$I=(\alpha,\ \beta),\ -\infty\leq\alpha<\beta\leq+\infty$, and its
moments
$$m_{k}=\int_{I}x^{k}f(x)d x,\ \ \ k=0,1,2,\ldots \eqno (3.1)$$
exist. Now consider the following deformation of moments with a
parameter $\lambda > 0$,
$$M_{k}=\frac{1}{1+\lambda k}\ m_{k}, \ \ \ \  k=0,1,2,\ldots . \eqno (3.2)$$

Question: When is there a unique density function $\sigma(x)$ which
is a solution of the moment problem of (3.2), i.e.
$$\int_{-\infty}^{+\infty}x^{k}\sigma(x)d x=M_{k},\ \ \ k=0,1,2,\ldots, \eqno (3.3)$$
and further how to determine $\sigma(x)$ from the given density
$f(x)$?

Note that two density functions $f(x)=g(x)$ are said  to be equal if
their distribution functions $\displaystyle\int _{-\infty}^{x}f(s)d
s$ and $\displaystyle\int _{-\infty}^{x}g(s) d s$ are equal at their
points of continuity. To make sure that the density is unique we
assume the moments $m_{k}, k=0,1,2,\ldots $ satisfy Carleman's
condition
$$\sum_{k=0}^{\infty} m_{2k}^{-1/2k}=\infty. \eqno (3.4)$$
It is obvious that the moments $M_{k}$ also satisfy Carleman's
condition
$$\sum_{k=0}^{\infty} M_{2k}^{-1/2k}=\infty. \eqno (3.5)$$
Due to Carleman's famous theorem (see [ST] or [Fe]) (3.5) assures
that the moment problem of (3.2) is determined, i.e. numbers $M_{k}$
determine a unique density whose $k$th  moment is $M_{k}$.

To prove that the moment problem of (3.2) has a solution $\sigma(x)$
whose spectrum support supp$\sigma(x)$ is to be contained in the
interval $J$, given in advance, we first introduce an important
theorem. Let $P(u)$ be any polynomial in $u$,
$$P(u)=\sum_{k}x_{k}\ u^{k}\eqno (3.6)$$
where numbers $x_{k}$ are real constants. Introduce the functional
$\mu (P)$ defined by
$$\mu (P)=\sum_{k}M_{k} x_{k}. \eqno (3.7)$$

\vskip 0.5cm \noindent {\bf Theorem 3.1} ([ST], Theorem 1.1) \hskip
0.2true cm A necessary and sufficient condition that the moment
problem defined by the sequence of moments $M_{k}$ shall have a
solution on $J$ is that the functional $\mu (P)$  be non-negative,
that is
$$\mu (P) \geq 0,\ \ \ \ \mbox{whenever}\ \ \ P(u) \geq 0\ \ \ \mbox{on}\ \ J. \eqno (3.8)$$

Note that Theorem 3.1 can be applied to derive  explicit necessary
and sufficient conditions  by a special choice of $J$, which depend
on representations of non-negative polynomials on $J$. In
particular, it is a well-known fact (see [P\'{o}S]) any polynomial
$P(u) \geq 0$ for all real $u$ can be presented as
$$P(u)=P_{1}(u)^{2}+P_{2}(u)^{2} \eqno (3.9)$$
where $P_{1}(u)$ and $P_{2}(u)$ are polynomials with real
coefficients.

If we take for $P(u)$ the particular polynomial
$P(u)=(x_{0}+x_{1}u+\cdots +x_{n}u^{n})^{2}$, we have
$$\mu (P)=\sum_{j, k=0}^{n}M_{j+k}x_{j}x_{k}. \eqno (3.10)$$
From the theory of quadratic forms and (3.9) it is well known that
the conditions of (3.8) are equivalent to
$$|M_{j+k}|_{j,k=0}^{n}> 0,\ \ n=0, 1,2, \cdots\eqno (3.11)$$
if the spectrum of the solution is not reducible to a finite set of
points.

\vskip .3cm
 Now we can state our theorem as follows:
 \vskip 0.5cm \noindent {\bf Theorem
3.2}  \hskip 0.2true cm A probability density $\sigma(x)$ on the
real axis with its $k\mbox{th}$ moment $M_{k}$ of (3.2) exists, and
is uniquely determined by the differential equation of first order
$$\sigma(x)-\lambda [x\sigma(x)]^{(1)}=f(x)\chi_{I} \eqno (3.12)$$
with the following conditions
$$\sigma(x)\geq 0,  \int_{-\infty}^{+\infty}\sigma(x) d x=1.  \eqno (3.13)$$

Proof. It is sufficient to prove (3.11) if  $(M_{j+k})_{j,k=0}^{n}$
is a  positive definite matrix. Taking $f(x)\chi_{I}$ for the
solution of the moments $m_{k}$ on the real axis, we have
$$|m_{j+k}|_{j,k=0}^{n}> 0,\ \ n=0, 1,2, \cdots ,\eqno (3.14)$$
that is,
$$\Delta_{n}=(m_{j+k})_{j,k=0}^{n} \ \mbox{is a  positive definite matrix} ,\ \ n=0, 1,2, \cdots. \eqno (3.15)$$

The formula of (3.2) shows that
 $$(M_{j+k})_{j,k=0}^{n}=\Delta_{n}*\Lambda_{n}\eqno (3.16)$$
 where $*$ represents  Schur product and
$$\Lambda_{n}=\big(\frac{1}{1+\lambda (j+k)}\big)_{j,k=0}^{n} ,\ \ n=0, 1,2, \cdots. \eqno (3.17)$$
Thus by the property of Schur product it is sufficient to prove that
$\Lambda_{n}$ is a positive definite matrix. Note that
$$\frac{1}{1+\lambda k}=\int_{0}^{1} t^{\lambda k} d t, \eqno (3.18)$$
then one can obtain
$$\mbox{det} \Lambda_{n}=\frac{1}{(n+1)!}\displaystyle\int_{0}^{1} \cdots \displaystyle\int_{0}^{1}
\displaystyle\prod_{0\leq j<k \leq n}
\big(t_{j}^{\lambda}-t_{k}^{\lambda}\big)^{2}
\displaystyle\prod_{j=0}^{n}d t_{j}>0. \eqno(3.19)$$

To derive the equation of (3.12) we make a Fourier transform and
write
$$H(t)= \int_{-\infty}^{+\infty}e^{i t x} \sigma(x) d x,\ \ \ F(t)= \int_{I}e^{i t x}f(x) d x. \eqno (3.20)$$
Thus, {\begin{eqnarray*}\hspace*{2.5cm}
  \int_{-\infty}^{+\infty}i t x \  e^{i t x} \sigma(x) d x
 \hspace*{-1.5mm}&=&\hspace*{-1.5mm}
 \sum\limits_{k=0}^{\infty}\frac{(i t)^{k+1}}{k!}M_{k+1}\\
  \hspace*{-1.5mm}&=&\hspace*{-1.5mm}
  \sum\limits_{k=0}^{\infty}\frac{(i t)^{k+1}}{(k+1)!}(k+1) M_{k+1}\\
\hspace*{-1.5mm}&=&\hspace*{-1.5mm}
\sum\limits_{k=0}^{\infty}\frac{(i t)^{k}}{k!}k
M_{k}.\hspace*{2.4cm} \ \ \ \ \ \ \ (3.21)
 \end{eqnarray*}}
Obviously,
 {\begin{eqnarray*}\hspace*{2.0cm}
  \int_{-\infty}^{+\infty}(1+\lambda i t x) \  e^{i t x} \sigma(x) d x
 \hspace*{-1.5mm}&=&\hspace*{-1.5mm}
 \sum\limits_{k=0}^{\infty}\frac{(i t)^{k}}{k!}(1+\lambda k)M_{k}\\
  \hspace*{-1.5mm}&=&\hspace*{-1.5mm}
  \sum\limits_{k=0}^{\infty}\frac{(i t)^{k}}{k!}m_{k}\\
\hspace*{-1.5mm}&=&\hspace*{-1.5mm} \int_{I}e^{i t x}f(x)d
x.\hspace*{2.5cm} \ \ \ (3.22)
 \end{eqnarray*}}
Namely,
$$H(t)+\lambda t H^{\ '}(t)=F(t). \eqno (3.23)$$

We also make an inverse Fourier transform. Combining
{\begin{eqnarray*}\hspace*{1.6cm}
  \frac{1}{2 \pi}\int_{-\infty}^{+\infty} \  e^{-i t x} t H^{\ '}(t)
  d t  \hspace*{-1.5mm}&=&\hspace*{-1.5mm}
  -\frac{1}{2 \pi i}\Big[\int_{-\infty}^{+\infty} \  e^{-i t x}  H^{\ '}(t) d t\Big]^{(1)}\\
  \hspace*{-1.5mm}&=&\hspace*{-1.5mm}
  -\frac{1}{2 \pi i}\Big[i x \int_{-\infty}^{+\infty} \  e^{-i t x}  H(t) d t \Big]^{(1)}\\
\hspace*{-1.5mm}&=&\hspace*{-1.5mm} -\big[x \sigma(x)
\big]^{(1)}\hspace*{3.5cm} \ \ \ (3.24)
 \end{eqnarray*}}
 and (3.23)
one obtains the differential equation on real axis
$$\sigma(x)-\lambda \big[x\sigma(x)\big]^{(1)}=f(x) \chi_{I} \eqno (3.25)$$
whose possible singular points are $0$, $\alpha$ and $ \beta$\ (if
$\alpha$ and $\beta$ are finite numbers). $\square$

\vskip .3cm
 Note that if $\alpha$ or $\beta$ is finite we have an
exact information about the spectrum support $J$ of $\sigma(x)$.

\vskip .3cm
 Case 1. $-\infty < \alpha \leq 0,\  \beta=+\infty$ (or $
\alpha =-\infty, 0 \leq \beta<+\infty$ similarly ). $J=I.$ \vskip
.3cm

Any non-negative polynomial $P(u)$ on $I=(\alpha, +\infty)$ can be
represented by
$$P(u)=P_{1}(u)^{2}+P_{2}(u)^{2} +(u-\alpha)(P_{3}(u)^{2}+P_{4}(u)^{2})\eqno (3.26)$$
where $P_{1}(u), P_{2}(u),P_{3}(u)$ and  $P_{4}(u)$ are polynomials
with real coefficients.

An analogous procedure to the above arguments of (3.9) --- (3.11)
one obtains the conditions of (3.8) are equivalent to the matrices
$$(M_{j+k})_{j,k=0}^{n},\ \ n=0, 1,2, \cdots\eqno (3.27)$$
and
$$(M_{j+k+1}-\alpha M_{j+k})_{j,k=0}^{n},\ \ n=0, 1,2, \cdots\eqno (3.28)$$
 are positive definite .

Note that
$$M_{k+1}-\alpha M_{k}=
\frac{m_{k+1}-\alpha m_{k}}{1+\lambda (k+1)}+(-\lambda
\alpha)\frac{m_{k}}{(1+ \lambda k)(1+\lambda (k+1))} \eqno (3.29)$$
and
$$(m_{j+k})_{j,k=0}^{n},(m_{j+k+1}-\alpha m_{j+k})_{j,k=0}^{n},\ \ n=0, 1,2, \cdots\eqno (3.30)$$
 are positive definite, by the property of Schur product one
 obtains the matrices of (3.27) and (3.28) are positive definite.

\vskip .3cm
 Case 2. $-\infty < \alpha \leq 0, 0<\beta<+\infty$ (or
$-\infty < \alpha <0, 0\leq\beta<+\infty$ similarly ). $J=I.$

\vskip .3cm

Any non-negative polynomial $P(u)$ on $I=(\alpha, \beta)$ can be
represented by
$$P(u)=P_{1}(u)^{2} +(u-\alpha)(\beta-u)P_{2}(u)^{2}\eqno (3.31)$$
where $P_{1}(u) \mbox{\ and\ } P_{2}(u)$ are polynomials with real
coefficients.

In this case the conditions of (3.8) are equivalent to the matrices
$$(M_{j+k})_{j,k=0}^{n},\ \ n=0, 1,2, \cdots\eqno (3.32)$$
and
$$(-M_{j+k+2}+(\beta-\alpha)M_{j+k+1}-\alpha \beta M_{j+k})_{j,k=0}^{n},\ \ n=0, 1,2, \cdots\eqno (3.33)$$
 are positive definite .

Note that $-M_{k+2}+(\beta-\alpha)M_{k+1}-\alpha \beta M_{k}$ can be
rewritten by {\begin{eqnarray*}
\frac{-m_{k+2}+(\beta-\alpha)m_{k+1}-\alpha \beta m_{k}}{1+\lambda
(k+1)}&\hspace*{-.2cm}+&\hspace*{-.2cm}(-\alpha
\beta)\frac{m_{k}}{(1+ \lambda k)(1+\lambda
(k+1))}\\
&\hspace*{-.2cm}+&\hspace*{-.2cm}\lambda \frac{m_{k+2}}{(1+ \lambda
(k+1))(1+\lambda (k+2))} \hspace*{1.5cm} \ (3.34)
\end{eqnarray*}}and
$$(m_{j+k})_{j,k=0}^{n},(-m_{k+2}+(\beta-\alpha)m_{k+1}-\alpha \beta m_{k})_{j,k=0}^{n},\ \ n=0, 1,2, \cdots\eqno (3.35)$$
 are positive definite, again by the property of Schur product one
 obtains the matrices of (3.32) and (3.33) are positive definite.

\vskip .3cm
 Remark 3.1 \ If $\alpha \beta>0$, we cannot make sure that the matrices of (3.33)
 are positive, e.g., $\alpha=1, \beta=2, n=0$
 $$-M_{2}+(\beta-\alpha)M_{1}-\alpha \beta M_{0}=-\frac{m_{2}}{1+2 \lambda}+\frac{m_{1}}{1+ \lambda}-2 m_{0}< 0 \eqno (3.36)$$
for the sufficiently large $\lambda$.

\vskip .3cm
 Case 3. $0 < \alpha <\beta$ (or $\alpha <\beta<0$
similarly ). $J=(0, \beta)$ (or $J=(\alpha, 0)$).

\vskip .3cm
 We take $f(x)$ for the density function on $(0, \beta)$
\[\hspace*{3.0cm}
\widetilde{f}(x)=\left\{
\begin{array}{ll}
f(x)\ \ \ \ \ \ \ \ \alpha<x<\beta\\
0\ \ \ \ \ \ \ \ \ \  \ \ \ 0 <x\leq \alpha \ \ \ \
\hspace*{3.5cm}(3.37)
\end{array}
\right.
\]
and by using Case 1 and Case 2 it is obvious.

\vskip .3cm
 Anyway, we get

 \vskip 0.5cm \noindent {\bf Theorem
3.3}  \hskip 0.2true cm A probability density $\sigma(x)$ with its
$k\mbox{th}$ moment $M_{k}$ of (3.2) exists, and is uniquely
determined by the following differential equation

(1) $ \alpha \leq 0 \leq\beta$,$$\sigma(x)-\lambda
[x\sigma(x)]^{(1)}=f(x),\ \  x\in I;\eqno (3.38)$$

(2) $ \alpha <\beta<0$,$$\sigma(x)-\lambda [x\sigma(x)]^{(1)}=f(x)
\chi_{I},\ \  x \in (\alpha, 0);\eqno (3.39)$$

(3) $ 0 <\alpha <\beta$,$$\sigma(x)-\lambda [x\sigma(x)]^{(1)}=f(x)
\chi_{I},\ \  x \in (0, \beta).\eqno (3.40)$$

\section{Proofs of Theorems 1.2 and 1.5}

Proof of Theorem 1.2.

In [Wig1] Wigner made use of the integral representation of Bessel
function of order 1 (pointed out by W.Feller to him) to get his
semicircle law. However, in the original analysis he got the
semicircle law by leading a differential equation. In the following
we will combine these two kinds of method to derive the density.

\vskip .3cm
 Uniqueness: Note that using (1.33) $\sum\limits_{j=0}^{k}C_{2k}^{j}
C_{2k-j}^{j} \leq 3^{2k}$ holds.  Writing $B=a+|b|$, then we have
$$M_{2k}\leq \sum_{j=0}^{k}C_{2k}^{j} C_{2k-j}^{j} a^{2j} b^{2k-2j}\leq (3B)^{2k}. \eqno (4.1)$$
Thus, the Carleman's condition is satisfied and the density function
is determined by the moments.

Derivation of the differential equation: Putting $2 a=1$, using the
integral representation of Bessel function of order zero (see [Sze]
) and calculating directly

{\begin{eqnarray*}\hspace*{1.3cm}
  \int_{-\infty}^{+\infty}(1+\lambda i t x) \  e^{i t x} \sigma(x) d x
 \hspace*{-1.5mm}&=&\hspace*{-1.5mm}
 \sum\limits_{k=0}^{\infty}\frac{(i t)^{k}}{k!}(1+\lambda k)M_{k}\\
  \hspace*{-1.5mm}&=&\hspace*{-1.5mm}
  \sum\limits_{k=0}^{\infty}\frac{(i t)^{k}}{k!}\sum_{j=0}^{[k/2]}C_{k}^{j} C_{k-j}^{j} a^{2j} b^{k-2j}\\
  \hspace*{-1.5mm}&=&\hspace*{-1.5mm}
  \sum\limits_{k=0}^{\infty}\frac{(i t)^{k}}{k!}L_{0}\Big (a
z+\frac{a}{z}+b\Big)^{k}\\
\hspace*{-1.5mm}&=&\hspace*{-1.5mm}
  L_{0}\ \sum\limits_{k=0}^{\infty}\frac{(i t)^{k}}{k!}\Big (a
z+\frac{a}{z}+b\Big)^{k}\\
\hspace*{-1.5mm}&=&\hspace*{-1.5mm}
  L_{0}\ \exp\Big (i t \big( a
z+\frac{a}{z}+b \big)\Big)\\
\hspace*{-1.5mm}&=&\hspace*{-1.5mm}
  e^{i t b} L_{0}\ \exp\Big (i t \big( a
z+\frac{a}{z} \big)\Big)\\
\hspace*{-1.5mm}&=&\hspace*{-1.5mm}
  e^{i t b} \ \sum\limits_{k=0}^{\infty}\frac{(i t a)^{2k}}{(2k)!}C_{2k}^{k} \\
\hspace*{-1.5mm}&=&\hspace*{-1.5mm} e^{i t b} \ J_{0}(2 a
t)\\
\hspace*{-1.5mm}&=&\hspace*{-1.5mm}
  e^{i t b} \frac{1}{\pi} \int_{-1}^{1}\ \frac{e^{i t x}}{\sqrt{1-x^{2}}}d x\\
  \hspace*{-1.5mm}&=&\hspace*{-1.5mm}
   \int_{b-1}^{b+1}\ e^{i t x}f_{b} (x)d x
  \hspace*{2.4cm} \ \ \ \ \ \ (4.2)
 \end{eqnarray*}}where
$$f_{b} (x)=\frac{1}{\pi}\   \frac{1}{\sqrt{1-(x-b)^{2}}}. \eqno (4.3)$$

By using Theorems 3.2 and 3.3 in Section 3 one obtains
$$\sigma(x)-\lambda \big[x\sigma(x)\big]^{(1)}=f_{b}(x)\chi_{I_{b}} \eqno (4.4)$$
and $\mbox{supp}(\sigma) = [B_{1}, B_{2}]$ for $\lambda >0$ while
 for $\lambda =0$, obviously
$$\sigma (x)=\frac{1}{\pi} \   \frac{1}{\sqrt{1-(x-b)^{2}}}.\eqno (4.5)$$

\vskip .3cm
 Now we give an exact solution of the equation of (4.4).
Note that when $\lambda >0$ the equation of (4.4) is known as a
Cauchy-Euler equation of order 1 (see [GN], P99). Thus we have

\vskip .2cm
 (1) $-1<b <1$. 0 is a singular point of the equation of
(4.4), and we have to determine how the solutions for $x<0$ and
$x>0$ can be pieced together to give solutions valid on the whole
interval $I_{b}$ (see [GN], P22 ). So one obtains
\[
\sigma(x)=\left\{
\begin{array}{ll}
 \   \displaystyle{\frac{1}{\lambda}x^{\frac{1}{\lambda}-1}\int_{x}^{b+1}\ s^{-\frac{1}{\lambda}} f_{b}(s) d s}\ \hspace*{1.8cm} x>0,\\
 \   \displaystyle{\frac{1}{\lambda}(-x)^{\frac{1}{\lambda}-1}\int_{b-1}^{x}\
(-s)^{-\frac{1}{\lambda}} f_{b}(s) d s}\ \ \ \ \ \ \  x<0.
\end{array}
\right. \eqno (4.6)
\]

Note that $x \sigma (x)$ is absolutely continuous on $I_{b}$. Thus
$\displaystyle\int_{b-1}^{b+1} \sigma (x) d x=1$ from the equation
of (4.4) by using $\sigma (b-1)=\sigma (b+1)=0$. Besides, $\sigma
(x)$ is continuous on $I_{b}$ when $0<\lambda<1$, while 0 is a
singular point  of $\sigma (x)$ when $\lambda \geq 1$.

\vskip .2cm
 (2) $b=\pm 1$.

If $b=1$, then
$$\sigma(x)=\frac{1}{\lambda}x^{\frac{1}{\lambda}-1}\int_{x}^{2}\ s^{-\frac{1}{\lambda}} \frac{1}{\sqrt{s(2-s)}}d s,
 \hspace*{1.3cm}\ x \in (0, 2).\eqno (4.7)$$
While $b=-1$, we have
$$\sigma(x)=\frac{1}{\lambda}(-x)^{\frac{1}{\lambda}-1}\int_{-2}^{x}\
(-s)^{-\frac{1}{\lambda}} \frac{1}{\sqrt{-s(s+2)}}d s,
\hspace*{1.3cm} x\in (-2, 0). \eqno (4.8)$$

\vskip .2cm
 (3) $b>1$. The equation of (4.4) can be rewritten by one
nonhomogeneous linear equation on $I_{b}=(b-1,\ b+1)$
$$\sigma(x)-\lambda \big[x\sigma(x)\big]^{(1)}=\frac{1}{\sqrt{1-(x-b)^{2}}} \eqno (4.9)$$
and the other homogeneous linear equation on $(0,\ b-1)$
$$\sigma(x)-\lambda \big[x\sigma(x)\big]^{(1)}=0. \eqno (4.10)$$

Next we solve the related nonhomogeneous linear equation of (4.9)
$$\sigma(x)=\frac{1}{\lambda}x^{\frac{1}{\lambda}-1}\int_{x}^{b+1}\ s^{-\frac{1}{\lambda}} f_{b}(s)d s \eqno (4.11)$$
and the related homogeneous linear equation of (4.10)
$$\sigma(x)=C_{+}x^{\frac{1}{\lambda}-1} \eqno (4.12)$$
where
$$C_{+}=\frac{1}{\lambda}\int_{b-1}^{b+1}\ s^{-\frac{1}{\lambda}} f_{b}(s)d s .\eqno (4.13)$$
Note that $C_{+}$ assures that $\sigma (x)$ is continuous on $(0,\
b+1)$ and $\int_{0}^{b+1} \sigma (x) d x=1$.

\vskip .2cm
 (4) $b<-1$. Similar to the case where $b>1$. One obtains
the solutions of the equation of (4.4) on $I_{b}=(b-1,\ b+1)$
$$\sigma(x)=\frac{1}{\lambda}(-x)^{\frac{1}{\lambda}-1}\int_{b-1}^{x}\
(-s)^{-\frac{1}{\lambda}} f_{b}(s)d s \eqno (4.14)$$ and on $(b+1,\
0)$
$$\sigma(x)= C_{-} (-x)^{\frac{1}{\lambda}-1} \eqno (4.15)$$ where
$$C_{-}=\frac{1}{\lambda}\int_{b-1}^{b+1}\ (-s)^{-\frac{1}{\lambda}} f_{b}(s)d s .\eqno (4.16)$$

\vskip .2cm
 In the end we will complete the proof of Theorem 1.5. First,
setting
$$\mathcal{H}_{n}=\mbox{span}\{p_{0}(x), p_{1}(x), \cdots, p_{n-1}(x)\},\eqno (4.17)$$
then $\mathcal{H}_{n}$ is an n-dimensional subspace of
$L^{2}(\omega)$. It is obvious that $\hat{p}_{0}(x)p(x),\cdots,$
$\hat{p}_{n-l-1}(x)p(x) $ is a family of normalized orthogonal
vectors in $\mathcal{H}_{n}$, extended by
$$e_{0}^{(n)}(x),\cdots, e_{l-1}^{(n)}(x),\hat{p}_{0}(x)p(x),\cdots,\hat{p}_{n-l-1}(x)p(x) \eqno (4.18)$$
to a normalized orthogonal base of $\mathcal{H}_{n}$ .

Let $P_{n}$ be a projective operator from $L^{2}(\omega)$ to
$\mathcal{H}_{n}$. We construct an  operator  from $\mathcal{H}_{n}$
to itself as follows,
$$T_{n}^{(k)}=P_{n}\circ A_{x}^{k}:\mathcal{H}_{n} \longrightarrow \mathcal{H}_{n} \eqno (4.19)$$
where $A_{x}$ is the multiplication by x.

\vskip 0.5cm \noindent {\bf Lemma  4.1} \hskip 0.2true cm Denote the
$k\mbox{th}$ moments of $\hat{\sigma}_{N}(x)$ and $\sigma_{N}(x)$ by
$M_{k}^{(N)}$ and $\hat{M}_{k}^{(N)}$ respectively, then
$$\hat{M}_{k}^{(N)}= M_{k}^{(N)}+\Theta_{N}
\eqno (4.20)$$ where
$$\Theta_{N}=\frac{1}{N\ (c_{N})^{k}}\Big (\sum_{j=N-l}^{N-1}\big\langle
A_{x}^{k}(\hat{p}_{j}p),\hat{p}_{j}p\big\rangle_{L^{2}(\omega)}-
\sum_{j=0}^{l-1}\big\langle
T_{N}^{(k)}(e_{j}^{(N)}),e_{j}^{(N)}\big\rangle_{L^{2}(\omega)}\Big
). \eqno(4.21)$$

\vskip 0.5cm \noindent {\bf Proof}. We first point out that
 {\begin{eqnarray*}
 \hspace*{3cm} M_{k}^{(N)}\hspace*{-1.5mm}&=&\hspace*{-1.5mm}\int x^{k}\sigma_{N}(x)d x\\
  \hspace*{-1.5mm}&=&\hspace*{-1.5mm}\frac{1}{N\ (c_{N})^{k}}\sum_{j=0}^{N-1}\int_{I} x^{k} p^{2}_{j}(x)\omega(x)d
  x\\
\hspace*{-1.5mm}&=&\hspace*{-1.5mm}\frac{1}{N\
(c_{N})^{k}}\sum_{j=0}^{N-1}\Big\langle x^{k}
p_{j},p_{j}\Big\rangle_{L^{2}(\omega)}\\
\hspace*{-1.5mm}&=&\hspace*{-1.5mm}\frac{1}{N\
(c_{N})^{k}}\sum_{j=0}^{N-1}\Big\langle T_{N}^{(k)}(
p_{j}),p_{j}\Big\rangle_{L^{2}(\omega)}\\
\hspace*{-1.5mm}&=&\hspace*{-1.5mm} \frac{\mbox{Tr}
\big(T_{N}^{(k)}\big)}{N\ (c_{N})^{k}}.\hspace*{5.0cm}\ \ \ \ \ \
(4.22)
 \end{eqnarray*}}On the other hand, by the normalized orthogonal base of (4.18) one
obtains
$$\mbox{Tr} \big(T_{N}^{(k)}\big)=\sum_{j=0}^{N-l-1}\big\langle
T_{N}^{(k)}(\hat{p}_{j}p),\hat{p}_{j}p\big\rangle_{L^{2}(\omega)}+
\sum_{j=0}^{l-1}\big\langle
T_{N}^{(k)}(e_{j}^{(N)}),e_{j}^{(N)}\big\rangle_{L^{2}(\omega)}.\eqno
(4.23)$$

Thus, {\begin{eqnarray*}
 \hspace*{-1cm} &&\hat{M}_{k}^{(N)}\hspace*{-1.5mm}=\hspace*{-1.5mm}\int x^{k}\hat{\sigma}_{N}(x)d x\\
  \hspace*{-1.5mm}&=&\hspace*{-1.5mm}\frac{1}{N\ (c_{N})^{k}}\sum_{j=0}^{N-1}\int_{I} x^{k} \hat{p}^{2}_{j}(x)p^{2}(x)\omega(x)d
  x\\
\hspace*{-1.5mm}&=&\hspace*{-1.5mm} \frac{1}{N\ (c_{N})^{k}}\Big
(\sum_{j=0}^{N-l-1}\big\langle \int_{I} x^{k}
\hat{p}^{2}_{j}(x)p^{2}(x)\omega(x)d x+ \sum_{j=0}^{l-1}\int_{I}
x^{k}
(e_{j}^{(N)}(x))^{2}\omega(x)d x \Big )+\Theta_{N}\\
\hspace*{-1.5mm}&=&\hspace*{-1.5mm} \frac{1}{N\ (c_{N})^{k}}\Big
(\sum_{j=0}^{N-l-1}\big\langle
T_{N}^{(k)}(\hat{p}_{j}p),\hat{p}_{j}p\big\rangle_{L^{2}(\omega)}+
\sum_{j=0}^{l-1}\big\langle
T_{N}^{(k)}(e_{j}^{(N)}),e_{j}^{(N)}\big\rangle_{L^{2}(\omega)}\Big )+\Theta_{N}\\
\hspace*{-1.5mm}&=&\hspace*{-1.5mm}M_{k}^{(N)}+\Theta_{N}.\hspace*{8.3cm}
\ \ \ \ \ \ \ (4.24)
 \end{eqnarray*}} $\square$

\vskip 0.5cm \noindent {\bf Lemma  4.2} \hskip 0.2true cm Write
$\|\cdot\|=\langle \cdot,\cdot\rangle^{1/2}_{L^{2}(\omega)}$,\ and
for $k=1,2,\cdots,$
$$\|A_{x}^{k}f\| \leq 3^{k}\Big(\prod_{j=0}^{k-1}D_{n+j}\Big) \|f\| ,\ \ \forall f\in \mathcal{H}_{n} \eqno (4.25)$$
where $$D_{n}=\max_{0\leq j \leq n}\{a_{j},\ |b_{j}|\}. \eqno
(4.26)$$

\vskip 0.5cm \noindent {\bf Proof}. Set
$f(x)=\sum\limits_{j=0}^{N-1}l_{j}p_{j}(x)$ and note that
 {\begin{eqnarray*}
 \hspace*{2cm} \|A_{x}f\|^{2}\hspace*{-1.5mm}&=&\hspace*{-1.5mm}\|\sum_{j=0}^{N-1}l_{j} x p_{j}\|^{2}\\
  \hspace*{-1.5mm}&=&\hspace*{-1.5mm}\|\sum_{j=0}^{N-1}l_{j}( a_{j+1}p_{j+1}+b_{j}p_{j}+a_{j}p_{j-1})\|^{2}\\
\hspace*{-1.5mm}&=&\hspace*{-1.5mm}\|\sum_{j}( l_{j-1}a_{j}+l_{j}b_{j}+l_{j+1}a_{j+1})p_{j}\|^{2}\\
\hspace*{-1.5mm}&=&\hspace*{-1.5mm}\sum_{j}|l_{j-1}a_{j}+l_{j}b_{j}+l_{j+1}a_{j+1}|^{2}\\
\hspace*{-1.5mm}&\leq &\hspace*{-1.5mm}3 D_{n}^{2}\sum_{j}
(l_{j-1}^{2}+l_{j}^{2}+l_{j+1}^{2})\\
\hspace*{-1.5mm}&\leq &\hspace*{-1.5mm} 9 D_{n}^{2}
\|f\|^{2}.\hspace*{5.5cm}\ \ \ \ \ \ \ (4.27)
 \end{eqnarray*}}Thus, $\|A_{x}f\|\leq 3 D_{n}\|f\| $  and (4.25) is easily
 proved. $\square$

\vskip .3cm
 Proof of Theorem 1.5: By (1.21) and Lemma 4.2, there
exist constants $C_{1}$ and $C_{2}$ which only depend on $k$ such
that
$$\big |\big\langle
A_{x}^{k}(\hat{p}_{j}p),\hat{p}_{j}p\big\rangle_{L^{2}(\omega)}\big
| \leq \|A_{x}^{k}(\hat{p}_{j}p)\|\leq 3^{k} D_{N+l+k-2}^{k}\leq
C_{1}
 (c_{N})^{k} \eqno (4.28)$$
and
$$\big |\big\langle
T_{N}^{(k)}(e_{j}^{(N)}),e_{j}^{(N)}\big\rangle_{L^{2}(\omega)}\big
| \leq \|P_{N}\| \ \|A_{x}^{k}e_{j}^{(N)}\|\leq 3^{k}
D_{N+k-1}^{k}\leq C_{2}
 (c_{N})^{k}. \eqno (4.29)$$
Thus, using (4.21), for the large $N$,
$$|\Theta_{N}|\leq \frac{1}{N\ (c_{N})^{k}}\Big (\sum_{j=N-l}^{N-1}C_{1}
 (c_{N})^{k}+
\sum_{j=0}^{l-1}C_{2}
 (c_{N})^{k}\Big
) =\frac{(C_{1}+C_{2})l}{N}.\eqno(4.30)$$

Again by Lemma 4.1 we get
$$\lim_{n\rightarrow \infty} \hat{M}_{k}^{(N)}=\lim_{n\rightarrow \infty} M_{k}^{(N)}
=M_{k}. \eqno (4.31)$$
 $\square$

\section*{Acknowledgements}
The authors would like to thank Professors Y. Chen, D. S. Lubinsky
and W. Van Assche  for helpful discussions on the relationship
between the density of eigenvalues and zero distribution of the
orthogonal polynomials. The first author is grateful to Professor Y.
Chen for his encouragement.

\bibliographystyle{amsunsrt}

\end{document}